\documentclass{emulateapj}

\newcommand{\dgr}{$^{\circ}$}
\newcommand{\etal}{et~al.\ }

\slugcomment{To appear in The Astrophysical Journal Letters}

\shorttitle{{\it Spitzer\/} Detections of Circumstellar Disks}
\shortauthors{Smith \etal}

\begin{document}

\title{{\it Spitzer\/} Far-Infrared Detections of Cold Circumstellar Disks}
\vskip 0.2in

\author{P. S. Smith\altaffilmark{1}, D. C. Hines\altaffilmark{2},
F. J. Low \altaffilmark{1},
R. D. Gehrz\altaffilmark{3},
E. F. Polomski\altaffilmark{3}, \& C. E. Woodward\altaffilmark{3}}
\altaffiltext{1}{Steward Observatory, The University of Arizona,
    Tucson, AZ 85721; psmith, flow@as.arizona.edu}

\altaffiltext{2}{Space Science Institute, 3100 Marine Street, Suite
A353, Boulder, CO 80303; dean.hines@colorado.edu}

\altaffiltext{3}{University of Minnesota, Department of Astronomy,
Minneapolis, MN 55455; gehrz, elwood, chelsea@astro.umn.edu}

\begin{abstract}

Observations at 70~$\mu$m with the {\it Spitzer Space Telescope\/}
have detected several stellar systems within 65~pc of
the Sun.  Of 18 presumably young systems detected in this study,
as many as 15 have 70~$\mu$m
emission in excess of that expected from their stellar photospheres.
Five of the systems with excesses are members of the Tucanae
Association.  The 70~$\mu$m excesses range from a factor of $\sim$2 to
nearly 30 times the expected photospheric emission from these stars.
In contrast to the 70~$\mu$m properties of these systems, there is
evidence for an emission excess at 24~$\mu$m for only HD~3003, confirming
previous results for this star.
The lack of a strong 24~$\mu$m excess in most of these systems
suggests that the circumstellar dust producing the IR excesses is
relatively cool ($T_{\rm dust}\lesssim 150$~K) and that there is
little IR-emitting material within the inner few AU of the primary
stars. 
Many of these systems lie close
enough to Earth that the distribution of the dust producing the IR excesses
might be imaged in scattered light at optical and near-IR
wavelengths.

\end{abstract}


\section{Introduction}

One of the major scientific objectives of the {\it Spitzer Space
Telescope\/} \citep[{\it Spitzer\/};][]{werner04} is to identify and
characterize circumstellar dust around a broad range of stars, both in
spectral type and age.  In this regard, the unprecedented sensitivity
and 
spatial 
resolution of {\it Spitzer\/} has proved to be a dramatically
successful tool in the infrared study of planetary debris systems
\citep[see, e.g.,][]{meyer04,uchida04,gorlova04,jura04,stapelfeldt04,
sloan04,rieke05,beichman05a,beichman05b,chen05,su05,uzpen05,stauffer05,
low05,kim05,bryden06,hines06}.  These early
investigations using data from {\it Spitzer\/} have generally shown
that ``warm'' material resulting in a strong signature at 24~$\mu$m,
indicating the presence of dust within the region where terrestrial
planet formation can take place, is quite rare for stars over $\sim$10
million years (Myr) old.  Also, the
frequency of finding colder material around stars of any age is higher
than finding dust that gives rise to an emission excess at 24~$\mu$m.
For instance, a study of the
TW~Hya Association (TWA) by \citet{low05} found that only one of 15
objects not previously detected by the {\it Infrared Astronomical
Satellite\/} ({\sl IRAS\/}), TWA~7, shows evidence for excess emission
at 24~$\mu$m.  However, based on the 70~$\mu$m flux density
measurement of TWA~7, 
the 24~$\mu$m 
excess is consistent with emission from dust at $T = 80$~K located at
least $\sim$7~AU from the star. 

The mixture of objects within the TWA
having little or no 24~$\mu$m excess, with a few objects (e.g., TW~Hya and
HD~98800B) of the same
age having among the strongest warm excesses observed, is a powerful
example of how rapidly dust within $\sim$5--10~AU of the stellar
primary can be removed from these systems.  Given that the TWA is
estimated to have an age of only 8--10 Myr, it is clear that dust
destruction mechanisms near the stellar primary, such as planet
building and the Poynting-Robertson effect, can become dominant or are
possibly even completed early in the formation of these systems
\citep[see, e.g.,][and references therein]{strom93}.
In fact, recent 3.6--24~$\mu$m {\it Spitzer\/} results for more distant
stellar associations and clusters,
ranging in age from $\sim$2--12~Myr,
begin to chronicle the rapid evolution of the inner dust disk
\citep[e.g.,][]{hartmann05,sicilia06,lada06}.

We have used {\it
Spitzer\/} to survey 112 stars that are close enough to Earth that
their distances have been measured by the {\it Hipparcos\/} Satellite
and exhibit some evidence that they are younger than $\sim$40~Myr
old.  \citet{low05} summarize the general selection criteria for this
sample. 
In this {\it Letter\/} we identify 18 stars in the sample
that have been detected at 70~$\mu$m.
Three of the detections are consistent with the
level of emission expected from the stellar photosphere.  Most of the
remaining 15 systems show substantial IR excesses, identifying them as
containing circumstellar material with $T \lesssim 150$~K. 

\section{Observations and Results}

Photometry at 24~$\mu\/$m and 70~$\mu\/$m of 112 stars within $\sim$100~pc of the Sun
was acquired using the Multi-band Imaging Photometer for {\it
Spitzer\/} \citep[MIPS;][]{rieke04}. 
The observations and reductions
closely follow those described in \citet{low05}.  All of the
70~$\mu\/$m observations
were made using the wide-field, default scale observing
mode.  Since publication of the MIPS observations of the TWA, the
photometric calibration conversion factors for the MIPS
bandpasses have been updated.  For the current study, we
use these revised factors, which are: 1.05 ($\pm$0.04) $\times
10^{-3}$~mJy~arcsec$^{-2}$~{\it DU\/}$^{-1}$ and
16.5 ($\pm$1.2)~mJy~arcsec$^{-2}$~{\it DU\/}$^{-1}$
for the 24 ~$\mu\/$m and
70~$\mu\/$m bandpasses, respectively \citep{engelbracht06}, where 
1~mJy = 10$^{-26}$~erg~cm$^{-2}$~s$^{-1}$~Hz$^{-1}$ and 
a data unit ({\it DU\/}) is the instrumental output produced by
the MIPS ``Data Analysis Tool''
(DAT) reduction pipeline \citep{gordon05}.
Uncertainties quoted for the
MIPS photometry include both the measurement precision and the
uncertainties in the photometric calibration factors.

Nineteen of 112 stars in the sample have 70~$\mu\/$m
detections with signal-to-noise ratios ($S/N\/$)$ \ge 5$, where $S/N\/$
is the measurement precision defined by the signal
measured from the object divided by the noise (per pixel) in the
background annulus, scaled to the photometric aperture. 
The 24~$\mu\/$m and
70~$\mu\/$m flux density measurements
for 18 of these stars are summarized in Table~1.
The {\it Spitzer\/} observations of HD~181327, a suspected member of the 
Tucanae Association having an IR excess previously detected by
{\sl IRAS\/}, are reported in \citet{schneider06}.
Also listed in Table~1 is the 70~$\mu\/$m detection
$S/N\/$. 
All 18 stars have a detection $S/N > 100$ at 24~$\mu\/$m
and the listed
photometric uncertainties are
dominated by the uncertainty
in the flux calibration factor for this bandpass.

Figure~1 plots a color-color diagram of the stars that shows
the range in the strengths of IR emission at both 24~$\mu\/$m and 70~$\mu\/$m.
Except for the IR excess identified previously by {\sl IRAS\/} for
HD~3003 \citep{whitelock89}, there is
little evidence for excess IR emission at 24~$\mu\/$m for the sample.
For comparison, we include the previous results for
TWA~7 in Figure~1 \citep{low05}. 
This star exhibits a 24~$\mu\/$m
emission excess that accounts for $\sim$40\% of the total measured
24~$\mu\/$m flux density. 
Even an excess this small produces a much
larger $F_{24\mu{\rm m}} / F_{K_s}\/$ than is observed for all of the
stars included in Figure~1 except for HD~3003.
Indeed, the remaining 17 systems are
located comfortably between the high-temperature Rayleigh-Jeans limit
for $F_{24\mu{\rm m}} / F_{K_s}\/$ and the flux density ratio expected
from a stellar photosphere with $T_{eff} = 3000$~K.

There is a possibility at 70~$\mu\/$m that flux from extragalactic
sources also falls within the 70\arcsec-diameter circular photometric aperture,
thereby resulting in 
a false detection or an overestimation of the far-IR
emission from a stellar system.
Based on {\it Spitzer\/} 70~$\mu\/$m source counts \citep{dole04},
we estimate that for each star there is
$\sim 1$\% chance that an extragalactic source with 
$F_{70\mu{\rm m}} > 15$~mJy can be found within
the aperture.
Therefore, we can expect a few cases where the measurements of
stars in a sample of over 100 are confused with background sources.
In fact, there is evidence for at least some contribution by background
objects to the 
measured 70~$\mu\/$m flux densities for four stars:  HD~2884, HD~11507,
BD+40\dgr 2208, and HD~124498.
For both BD+40\dgr 2208 and HD~124498, there are two faint sources within 
20\arcsec\ of the stars in the 24~$\mu\/$m images that could contribute
to the total flux measured at 70~$\mu\/$m.
The displacement of the centroid of the 70~$\mu\/$m flux by about 10\arcsec\ 
(1~pixel) is consistent with the two background sources providing most of the
measured flux instead of BD+40\dgr 2208.
Although the centroid of the flux for HD~124498 is at the 
expected location on the 70~$\mu\/$m array, we cannot rule out the 
possibility that a fraction
of the signal is from the sources seen at 24~$\mu\/$m.
The centroid of HD~11507 is offset from the nominal 70~$\mu\/$m array location
by $3\times$ the $\sim 0.3$~pixel (rms) deviations observed relative to the 
stars' positions on the 24~$\mu\/$m detector.
In this case, the position angle and magnitude of the offset is consistent
with at least half of the 70~$\mu\/$m flux coming from a source 
apparent at 24~$\mu\/$m that is $\sim$20\arcsec\
away from HD~11507.
Likewise, the 70~$\mu\/$m source measured for HD~2884, is displaced from the
aperture center by about 1~pixel, but there is no corresponding
source seen at 24~$\mu\/$m.
The values listed in Table~1 for $F_{70\mu{\rm m}}\/$ for these four stars
can be considered upper limits because of the uncertain contribution by
possible background sources.

In contrast to the lack of strong 24~$\mu\/$m excesses in the
sample, only three stars (HD~2885, HD~98230, and HD~193924) have
measured 70~$\mu\/$m flux densities that are fully consistent with
emission from their photospheres.  The 11 objects that show no evidence
for contamination of their far-IR photometry by background sources
have 70~$\mu\/$m emission ranging
from $\sim$2--30 times greater than expected from a stellar
photosphere (Figure~1), implying thermal emission from dust residing in
circumstellar disks.
The broad-band IR spectra of three systems that show a representative range 
of emission strengths at 70~$\mu\/$m are displayed in Figure~2.

\section{Discussion}

Table~2 summarizes the upper limits for the temperature
($T_D\/$) and
minimum distance from the stellar primary ($R_D\/$) of the
circumstellar dust detected by {\it Spitzer\/} assuming blackbody 
dust grains in thermal equilibrium with the stellar radiation field.
Except where noted in the table, $T_D\/$ is an estimated color
temperature based on the flux ratio of the excess emission at
24~$\mu\/$m and 70~$\mu\/$m.  For most systems, the presence of dust
emission at 24~$\mu\/$m is uncertain and we have assigned an extreme
upper limit on the strength of any possible flux excess at this
wavelength by setting the photospheric flux density to the value
dictated by the Rayleigh-Jeans limit based on the 2MASS $K_s\/$-band
flux density [i.e., $F_{24\mu{\rm m}} =
(\lambda_{K_s}/\lambda_{24\mu{\rm m}})^2 F_{K_s}\/$; see Figure~1].
The upper limit chosen for the 24~$\mu\/$m excess leads to an upper
limit for the temperature of the emitting material, and in turn, a
lower limit for the distance from the star of the dust.

The fraction of 70~$\mu\/$m excess detections in the overall 
sample is $\sim$10--15\% (12--16 of 112 objects).
Included in our MIPS survey are
22 stars believed to be members of the Tucanae Association
\citep{zuckerman00,zuckerman01}.
Eight of these systems, including HD~181327 \citep{schneider06} and HD~2884
(see \S2), are detected at 70~$\mu\/$m, although
the measurements for HD~2885 and
HD~193924 are consistent with purely photospheric emission. 
The {\it Spitzer\/} observations also confirm the detection of HD~3003
at 25~$\mu\/$m by {\sl IRAS\/}, and this object
is the only member of
the Tucanae Association that shows emission from dust at $T_D >
200$~K as suggested by the {\it Spitzer\/} and
{\sl IRAS\/} 12--70~$\mu\/$m data (see Figure~2).

Comparison of the results for the $\sim$20--40~Myr Tucanae Association
with those for the younger $\sim$8--10~Myr TWA
shows about a three-fold
decrease in the fraction of detected 24~$\mu\/$m excesses for the
older systems.\footnote{Both associations have similar mean distance
from the Sun, $d \sim 40$--60~pc.}
The Tucanae Association has about the
same fraction of detected emission excesses at 70~$\mu\/$m as seen in
the TWA.
Cold material, however, is detected by {\it Spitzer\/} around
Tucanae systems having spectral types B--G, not the
late-type stars that also dominate the membership of the TWA.

Finally, we note that the relative proximity of these systems to the
Sun, together with the fact that the observed 70~$\mu\/$m
flux excesses reveal relatively cold material, imply that
it is possible to
image the distribution of the circumstellar dust in
scattered light at optical and near-IR wavelengths for many of these objects
with current high-resolution capabilities.  In the last column
of Table~2, we list the lower limit for the angular extent of the
IR-emitting dust.  For most systems, the dust is at least 0\farcs
1--0\farcs 2 from the stars.
Indeed, recent
{\it Hubble Space Telescope\/}
observations of HD~181327
using the NICMOS and ACS coronographs have
revealed the circumstellar disk in scattered optical and near-IR
light, with peak intensity at $\sim 1\farcs 7\/$ from the star
\citep{schneider06}.

\acknowledgments

This work is based on observations made with {\it Spitzer\/},
which is operated by the Jet Propulsion Laboratory
(JPL), California Institute of Technology (CIT), under National
Aeronautics and Space Administration (NASA) contract 1407.  We thank
NASA, JPL, and the {\it Spitzer\/} Science Center for support through
contracts 960785, 959969, and 1256424 to The University of Arizona and
contracts 1256406 and 1215746 to the University of Minnesota.
This research has made use of the {\sl SIMBAD\/} database, operated at
CDS, Strassbourg, France, and data products
from the 2MASS, which is a joint project of the University of
Massachusetts and the Infrared Processing and Analysis Center/CIT,
funded by NASA and the National Science Foundation.


\vspace{0.1in}
\begin{deluxetable}{lrcrrcrrr}
\tablecolumns{9}
\tablewidth{0pc}
\tabletypesize{\footnotesize}
\tablecaption{Eighteen Stellar Systems Detected at 70~Microns}
\tablehead{
\colhead{Star} &
\colhead{\hfil HIP\tablenotemark{a}} & \colhead{Spectral Type\tablenotemark{b}} &
\colhead{distance (pc)\tablenotemark{a}} &
\colhead{$V_T\/$\tablenotemark{a}} &
\colhead{$K_s\/$\tablenotemark{c}} &
\colhead{$F_{24\mu{\rm m}}$ (mJy)} &
\colhead{$F_{70\mu{\rm m}}$ (mJy)} &
\colhead{($S/N\/$)$_{70\mu{\rm m}}$} }

\startdata

HD~1466\tablenotemark{d} & 1481 & F8/G0V & 41\ $\pm$\ 1 & 7.53 & 6.15 & 32.9\ $\pm$\ 1.3 & 22.4\ $\pm$\ 3.3 & 7.7 \\
HD~2884\tablenotemark{d,}\tablenotemark{e} & 2484 & B9V & 43\ $\pm$\ 1 & 4.33 & 4.48 & 101.7\ $\pm$\ 8.6 & 56.8\ $\pm$\ 5.2\tablenotemark{f} & 17.2 \\
HD~2885\tablenotemark{d,}\tablenotemark{e} & 2487 & A2V & 53\ $\pm$\ 13 & 4.53 & 4.11 & 156.1\ $\pm$\ 4.5 & 18.5\ $\pm$\ 3.8 & 5.2 \\
HD~3003\tablenotemark{d,}\tablenotemark{e} & 2578 & A0V & 46\ $\pm$\ 1 & 5.08 & 4.99 & 223.9\ $\pm$\ 9.0 & 62.2\ $\pm$\ 5.1 & 22.8 \\
HD~11507 & 8768 & K7 & 11.1\ $\pm$\ 0.2 & 9.07 & 5.18 & 67.6\ $\pm$\ 2.7 & 24.2\ $\pm$\ 1.9\tablenotemark{f} & 26.6 \\
BD+40$^\circ$2208 & 46383 & K5 & 32\ $\pm$\ 2 & 9.99 & 6.62 & 18.7\ $\pm$\ 0.8 & 27.6\ $\pm$\ 3.0\tablenotemark{f} & 12.0 \\
HD~82443 & 46843 & K0V & 17.8\ $\pm$\ 0.3 & 7.16 & 5.12 & 72.3\ $\pm$\ 2.9 & 29.5\ $\pm$\ 4.4 & 7.7 \\
HD~84075 & 47135 & G2V & 63\ $\pm$\ 3 & 8.65 & 7.16 & 12.4\ $\pm$\ 0.5 & 34.4\ $\pm$\ 3.7 & 12.4 \\
HD~95650 & 53985 & M0 & 11.7\ $\pm$\ 0.2 & 9.83 & 5.69 & 44.6\ $\pm$\ 1.8 & 14.8\ $\pm$\ 2.9 & 5.5 \\
HD~98230\tablenotemark{e} & 55203 & F8.5V & 7.3\ $\pm$\ 0.9 & 3.79 & 2.14 & 968.5\ $\pm$\ 38.7 & 93.9\ $\pm$\ 10.0 & 12.5 \\
BD+21$^\circ$2486\tablenotemark{e} & 63942 & K5 & 18.6\ $\pm$\ 0.9 & 9.62 & 6.04 & 34.4\ $\pm$\ 1.4 & 44.7\ $\pm$\ 3.9 & 18.9 \\
HD~124498\tablenotemark{e} & 69562 & K4V & 26\ $\pm$\ 5 & 10.52 & 6.60 & 20.8\ $\pm$\ 0.8 & 118.4\ $\pm$\ 14.3\tablenotemark{f} & 44.0 \\
HD~177171\tablenotemark{d,}\tablenotemark{e} & 93815 & F7V & 52\ $\pm$\ 2 & 5.24 & 4.06 & 217.4\ $\pm$\ 8.7 & 45.1\ $\pm$\ 5.5 & 10.2 \\
HD~180134 & 94858 & F7V & 46\ $\pm$\ 1 & 6.42 & 5.10 & 63.8\ $\pm$\ 2.7 & 16.2\ $\pm$\ 3.4 & 5.1 \\
HD~192263\tablenotemark{e,}\tablenotemark{g} & 99711 & K2V & 19.9\ $\pm$\ 0.4 & 7.88 & 5.54 & 40.8\ $\pm$\ 1.6 & 28.7\ $\pm$\ 2.9 & 13.7 \\
HD~193924\tablenotemark{d,}\tablenotemark{e} & 100751 & B2IV & 56\ $\pm$\ 2 & 1.90 & 2.48 & 628.5\ $\pm$\ 25.1 & 63.5\ $\pm$\ 6.5 & 13.5 \\
HD~202917\tablenotemark{d} & 105388 & G5V & 46\ $\pm$\ 2 & 8.74 & 6.91 & 20.0\ $\pm$\ 0.8 & 35.1\ $\pm$\ 3.8 & 12.3 \\
HD~218340 & 114236 & G3V & 57\ $\pm$\ 2 & 8.52 & 6.95 & 11.6\ $\pm$\ 0.5 & 32.0\ $\pm$\ 3.5 & 11.7 \\
\enddata

\tablenotetext{a}{Hipparcos designations, distances, and visual {\it Tycho\/}
($V_T\/$) magnitudes
are from the {\it Hipparcos\/}
and {\it Tycho\/} Catalogs \citep{perryman97,hog00}.}
\tablenotetext{b}{Spectral type is quoted from the {\sl SIMBAD\/}
Astronomical Database.}
\tablenotetext{c}{Apparent $K_s\/$ magnitudes are from the Two-Micron All Sky
Survey (2MASS) Point Source
Catalog.}
\tablenotetext{d}{Probable or possible members of the Tucanae Association
\citep{zuckerman00}.}
\tablenotetext{e}{Unresolved double stars or spectroscopic binary systems.
In each case, the total magnitudes and flux densities are listed for the 
systems.}
\tablenotetext{f}{Possible flux from background sources included within
the photometry aperture.
These values should be considered upper limits to the 70~$\mu\/$m
flux density possibly detected from the stellar system.}
\tablenotetext{g}{Possible planetary system detected \citep{santos03}.}

\end{deluxetable}

\vspace{0.5in}
\begin{deluxetable}{lrcc}
\tablecolumns{4}
\tablewidth{20pc}
\tabletypesize{\small}
\tablecaption{Properties of the Circumstellar Dust}
\tablehead{
\colhead{Star}  &
\colhead{$T_{D}\/$ (K)\tablenotemark{a}} &
\colhead{$R_{D}\/$ (AU)} &
\colhead{$\theta_D\/$ ($\arcsec\/$)} }
\startdata

HD~1466 & $<$ 115 & $>$ 7.2 & $>$ 0.18 \\
HD~2884 & $<$ 90 & $>$ 65 & $>$ 1.5 \\
HD~3003\tablenotemark{b} & 230 & 6.7 & 0.14 \\
HD~11507 & $<$ 135 & $>$ 0.9 & $>$ 0.08 \\
BD+40$^\circ$2208 & $<$ 90 & $>$ 4.9 & $>$ 0.15 \\
HD~82443 & $<$ 130 & $>$ 3.2 & $>$ 0.18 \\
HD~84075 & $<$ 80 & $>$ 14 & $>$ 0.23 \\
HD~95650 & $<$ 150 & $>$ 0.5 & $>$ 0.04 \\
BD+21$^\circ$2486 & $<$ 95 & $>$ 3.3 & $>$ 0.18 \\
HD~124498 & $<$ 70 & $>$ 4.1 & $>$ 0.16 \\
HD~177171 & $<$ 220 & $>$ 7.3 & $>$ 0.14 \\
HD~180134 & $<$ 145 & $>$ 8.6 & $>$ 0.19 \\
HD~192263 & $<$ 90 & $>$ 5.2 & $>$ 0.26 \\
HD~202917 & $<$ 90 & $>$ 7.4 & $>$ 0.16 \\
HD~218340 & $<$ 70 & $>$ 17 & $>$ 0.31 \\

\enddata

\tablenotetext{a}{The limit for $T_D\/$ is calculated from the IR
excess measured at 70~$\mu$m and a conservative upper limit for the
excess at 24~$\mu$m (see text).}
\tablenotetext{b}{The listed
$T_D\/$ is based on a Planck function fit to the {\it Spitzer\/}
photometry corrected for the contribution by
the stellar photosphere.}

\end{deluxetable}

\clearpage

\begin{figure}
\figurenum{1}
\includegraphics{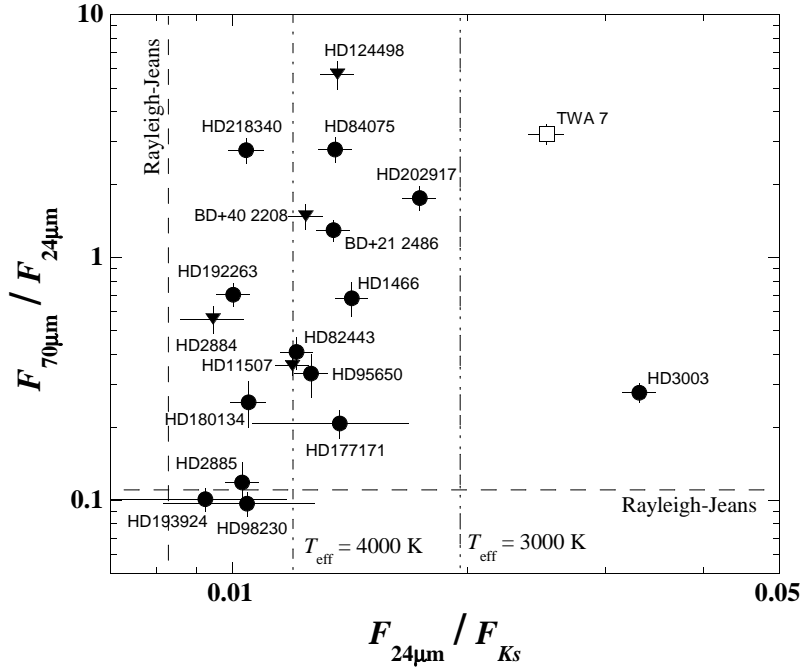}
\vspace{4.0in}
\caption{MIPS photometry compared to the 2MASS $K_s\/$-band
measurements of the stellar sample.
The dashed lines represent the $F_{24\mu\/{\rm m}\/}$-to-$F_{Ks}\/$
and $F_{70\mu\/{\rm m}\/}$-to-$F_{24\mu\/{\rm m}\/}$
flux density ratios in the limit where the photometric bands lie along
the Rayleigh-Jeans tail of the stellar spectrum.
This is a poor approximation in the case of $F_{24\mu\/{\rm m}\/} / F_{Ks}\/$
for most stars and, therefore, the ratios based on photospheric models for
$T_{eff} = 3000$~K and 4000~K are also shown.
{\it Triangles\/} represent upper limits to
$F_{70\mu\/{\rm m}\/}/F_{24\mu\/{\rm m}\/}$ for four stars that are likely 
to include flux from background sources in the measurements of
$F_{70\mu\/{\rm m}\/}$ (see \S2).}
\label{fig1}
\end{figure}


\begin{figure}
\figurenum{2}
\includegraphics{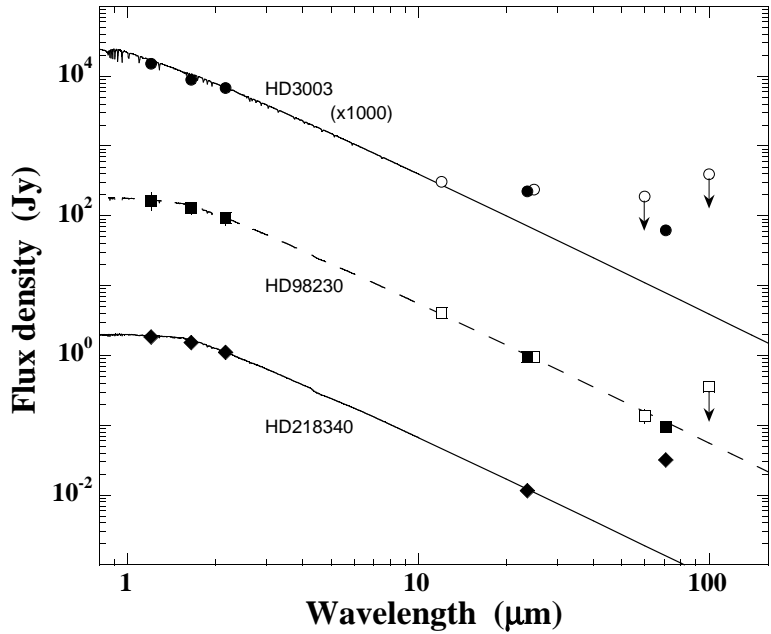}
\vspace{3.6in}
\caption{Broad-band IR spectra of three objects detected at 70~$\mu\/$m.
{\it Spitzer\/} and 2MASS photometry is represented by {\it filled\/} symbols,
and {\sl IRAS\/} measurements are shown as {\it open\/} symbols.
The {\sl IRAS\/} data are from the {\sl IRAS\/} Faint Source Catalog
\citep{moshir92} and are color corrected unless the displayed
point represents an upper
limit to the flux density ({\it arrows\/}).
Also shown is a photospheric model \citep{kurucz79} for each star that
is based on the
2MASS photometry and the stellar spectral type.
HD~218340 has one of the largest 70~$\mu\/$m excesses in the sample, whereas
the IR emission from HD~98230 is dominated by the stellar photosphere
out to 70~$\mu\/$m.
For clarity, the spectrum for HD~3003 has been scaled by a factor of 1000,
so for this object, the flux density is in units of milli-Janskys.}
\label{fig2}
\end{figure}
\end{document}